 \documentclass[final,5p,times,twocolumn]{elsarticle}

\usepackage{amssymb}

\usepackage{epsfig}
\usepackage{amsfonts}
\usepackage[bbgreekl]{mathbbol}
\usepackage{mathtools}
\usepackage{amsthm}
\usepackage{amsbsy}
\usepackage{multirow}
\usepackage{slashed}
\usepackage{esint}
\usepackage{epsfig}
\usepackage{amsmath,latexsym}
\usepackage[T1]{fontenc}
\usepackage{hyperref}
\usepackage{color}

\def\NO{\nonumber}

\newcommand{\be}{\begin{equation}}
	\newcommand{\ee}{\end{equation}}

\def\bea{\begin{eqnarray}}
	\def\eea{\end{eqnarray}}

\newcommand{\ca}{{\cal A}}

\newcommand{\ce}{{\cal E}}
\newcommand{\ck}{{\cal K}}

\newcommand{\cm}{{\cal M}}
\newcommand{\cl}{{\cal L}}
\newcommand{\cf}{{\cal F}}

\newcommand{\co}{{\cal O}}
\newcommand{\cp}{{\cal P}}
\newcommand{\cs}{{\cal S}}
\newcommand{\cq}{{\cal Q}}
\newcommand{\ct}{{\cal T}}

\def\a{\alpha}

\def\d{\delta}
\def\e{\epsilon}
\def\f{\phi}
\def\g{\gamma}

\def\j{\psi}

\def\m{\mu}
\def\n{\nu}

\def\om{\omega}
\def\p{\pi}

\def\th{\theta}
\def\r{\rho}
\def\s{\sigma}

\def\x{\xi}

\def\D{\Delta}

\def\P{\Pi}

\def\bs{\bbalpha}
\def\bs{\bbbeta}
\def\bs{\bbgamma}
\def\bs{\bbdelta}
\def\bs{\bbepsilon}
\def\bs{\bbeta}
\def\bs{\bbsigma}

\def\ca{{\cal A}}

\def\cc{{\cal C}}

\def\ce{{\cal E}}
\def\cf{{\cal F}}

\def\ch{{\cal H}}

\def\ck{{\cal K}}
\def\cl{{\cal L}}
\def\cm{{\cal M}}

\def\co{{\cal O}}
\def\cp{{\cal P}}
\def\cq{{\cal Q}}
\def\car{{\cal R}}
\def\cs{{\cal S}}
\def\ct{{\cal T}}

\def\bb#1{\ensuremath{\mathbb{#1}}} 

\def\bo{{\raise-.3ex\hbox{\large$\Box$}}}               
\def\pa{\partial}                                       
\def\face{{\raise.2ex\hbox{$\displaystyle \bigodot$}\mskip-2.2mu \llap {$\ddot\smile$}}}                                    
\def\>{\rangle}                                      
\def\<{\langle}                                      

\newcommand{\sub}[1]{{}_{(#1)}{}}    					 
\def\Hat#1{\widehat{#1}}                             
\def\leftrightarrowfill{$\mathsurround=0pt \mathord\leftarrow \mkern-6mu
	\cleaders\hbox{$\mkern-2mu \mathord- \mkern-2mu$}\hfill
	\mkern-6mu \mathord\rightarrow$}        
\def\dvec#1{\vbox{\ialign{##\crcr
			\leftrightarrowfill\crcr\noalign{\kern-1pt\nointerlineskip}
			$\hfil\displaystyle{#1}\hfil$\crcr}}}           

\def\-{\hphantom{-}}

\journal{Nuclear Physics B}

\begin{document}

\begin{frontmatter}




\title{\vskip-0.5in
	\begin{minipage}{6.5in}
		\begin{flushright}
			{\small IFT UAM/CSIC-14-110\\
				\small SISSA 55/2014/FISI}
		\end{flushright}
	\end{minipage}\\
	\vskip 0.5in 
	Hyperscaling violating Lifshitz holography}


\author{Ioannis Papadimitriou}
\ead{Ioannis.Papadimitriou@sissa.it}
 
\address{Instituto de F\'isica Te\'orica UAM/CSIC, Universidad Aut\'onoma de Madrid, Madrid 28049, Spain \\ 
	and \\
	SISSA and INFN - Sezione di Trieste,
	Via Bonomea 265, I 34136 Trieste, Italy
	}

\begin{abstract}
We present an overview of the construction of the general holographic dictionary for asymptotically locally Lifshitz and hyperscaling violating Lifshitz backgrounds with arbitrary dynamical exponents $z$ and $\th$, compatible with the null energy condition, which was recently developed in \cite{Chemissany:2014xpa,hvLf}. A concrete definition of asymptotically locally Lifshitz and hyperscaling violating Lifshitz backgrounds is provided in the context of a generic  bottom-up Einstein-Proca-Dilaton theory, and a systematic procedure for solving the radial Hamilton-Jacobi equation via a covariant expansion in eigenfunctions of two commuting operators is presented. The resulting asymptotic solution of the Hamilton-Jacobi equation is subsequently used to derive the full holographic dictionary, including the Fefferman-Graham asymptotic expansions and the non-relativistic holographic Ward identities.       

\end{abstract}

\begin{keyword}

AdS/CFT \sep AdS/CMT \sep holography \sep Lifshitz \sep hypesrcaling violation \sep holographic renormalization



\end{keyword}

\end{frontmatter}


\section{Introduction}
\label{intro}

Holographic techniques have emerged in recent years as a powerful and wide-ranging new tool in the study of quantum systems exhibiting strongly coupled dynamics. In condensed matter physics the gauge/gravity duality has been applied successfully  to -- among numerous other systems -- quantum critical points exhibiting Lifshitz \cite{Kachru:2008yh,Taylor:2008tg} and Schr\"odinger \cite{Balasubramanian:2008dm,Son:2008ye} symmetry. 
The invariance of these systems under non-relativistic scaling transformations constrains the form of the correlation functions and leads to definite scaling laws for the various physical observables, a property often referred to as `hyperscaling'.   

More recently, non-relativistic backgrounds with scale symmetry as a {\em conformal} isometry -- but not an isomentry -- were put forward as holographic duals to quantum systems exhibiting 
hyperscaling violation \cite{Charmousis:2010zz,Huijse:2011ef,Dong:2012se,Shaghoulian:2011aa}. Backgrounds with Lifshitz symmetry as the conformal isometry group are referred to as hyperscaling violating Lifshitz (hvLf) and are of the form 
\be\label{thetaz}
ds_{d+2}^2 =\frac{  du^2 - u^{-2(z-1)} dt^2 + d\vec x^2 }{\ell^{-2} u^{2(d-\theta)/d} },
\ee
where $d$ is the number of spatial dimensions, $z$ is the usual Lifshitz dynamical exponent and $\theta$ is known as the hyperscaling violation parameter. This reflects the fact that for $\th\neq 0$ the scaling transformation 
\be
\label{scalecovariance}
\vec x \to \lambda \vec x
,\;
t \to \lambda^{z} t
,\;
u \to \lambda u,
\ee
is only a conformal isometry of the metric (\ref{thetaz}). Understanding the physics of such backgrounds is aided by the observation that for $z=1$, in which case Poincar\'e invariance is restored, the metric (\ref{thetaz}) is identical to the metric of non-conformal branes, i.e. D$p$ branes with $p\neq 3$, in the Einstein frame and with $\th$ related to the spatial dimension $p$ \cite{Kanitscheider:2008kd,Wiseman:2008qa}. Moreover, as for non-conformal branes, non-relativistic hyperscaling violation requires the presence of a linear dilaton. Holography for the hvLf backgrounds (\ref{thetaz}) should therefore be related to that of Lifshitz backgrounds, obtained from (\ref{thetaz}) by setting $\th=0$, in the same way that holography for non-conformal branes is related to that of D$3$ branes \cite{Gherghetta:2001iv}. As for non-conformal branes this connection is best seen in the 
`dual frame' \cite{Kanitscheider:2008kd}, where the metric is asymptotically Lifshitz. The only difference between Lifshitz and hvLf backgrounds in this frame is the presence of a linear dilaton in the latter case. 

By going to the dual frame one can immediately identify the location of the ultraviolet (UV) regime of the dual quantum field theory as the limit $u\to 0$, independently of the values of the exponents $z$ and $\th$ \cite{hvLf}. The same conclusion can be reached by computing the energy of supergravity fluctuations around the background (\ref{thetaz}) \cite{Peet:1998wn,Dong:2012se} in any frame. Restrictions on the exponents $z$ and $\th$ are imposed, however, by the null energy condition. There are seven distinct solutions of the null energy condition \cite{hvLf}, all of which except for two require $z>1$. The two solutions with $z<1$ require $\th >d+z$, which in turn implies that the on-shell action is not UV divergent and hence there are no well defined Fefferman-Graham asymptotic expansions \cite{hvLf}. This is analogous to the case of D$6$ branes, where for the same reason there are no well defined UV expansions \cite{Kanitscheider:2008kd}. In the marginal case $\th=d+z$ the on-shell action diverges logarithmically and it would be interesting to study this case in more detail. As in \cite{hvLf}, here we focus on all solutions of the null energy condition with $z>1$ so that there are well defined UV asymptotic expansions.   

Lifshitz and hyperscaling violating Lifshitz backgrounds have been discussed in a number of recent papers, including \cite{Donos:2010tu,Gouteraux:2012yr,Gath:2012pg,Bueno:2012vx,Narayan:2012wn,Moroz:2014ska,Hartong:2014oma,Hartong:2014pma,Arav:2014goa} and references therein. Holographic renormalization for asymptotically Lifshitz backgrounds in the Einstein-Proca theory has been addressed directly in  \cite{Ross:2009ar,Horava:2009vy,Ross:2011gu,Mann:2011hg,Baggio:2011cp,Baggio:2011ha,Griffin:2011xs,Griffin:2012qx}, while in \cite{Cassani:2011sv,Chemissany:2011mb,Chemissany:2012du,Christensen:2013lma,Christensen:2013rfa,Korovin:2013bua,Korovin:2013nha} special cases of the holographic dictionary for Lifshitz backgrounds were obtained indirectly via suitable embeddings or limits of AdS backgrounds. The main differences in the approach of \cite{Chemissany:2014xpa,hvLf} relative to previous works on holographic renormalization of such backgrounds are: 
\begin{enumerate}
	
	\item[i)] The analysis includes a very large class of theories, parameterized by the arbitrary functions in the action (\ref{action}). 
	
	\item[ii)] Arbitrary values of the spatial dimension $d$ and of the exponents $z$ and $\th$ are considered, as long as they are consistent with the null energy condition.
	
	\item[iii)] We consider backgrounds which are asymptotically locally Lifshitz or hvLf in the far UV. 
	
	\item[iv)] We choose to work entirely in the metric formalism, avoiding the use of vielbeins. 
	
	\item[v)] The holographic dictionary is obtained systematically by recursively solving the radial Hamilton-Jacobi equation. 
	
\end{enumerate}      
The fact that we treat Lifshitz and hvLf backgrounds as UV complete allows us to identify the set of gauge-invariant local operators in the dual non-relativistic theory. Moreover, it facilitates a genuine computation of correlation functions in the hyperscaling violating theory valid at all energy scales, without the need to take the small momentum limit. The behavior of the correlation functions as functions of the momenta in the presence of hyperscaling violation is an interesting open question which we intend to revisit in future work. Another interesting open question that can be addressed using the results of \cite{Chemissany:2014xpa,hvLf} is the thermodynamics of asymptotically Lifshitz and hvLf black holes. Finally, although we choose to work in the metric formalism, our results are in agreement with previous work using the vielbein formalism, such as \cite{Ross:2011gu} in the case of the Einstein-Proca theory. We believe that the Newton-Cartan geometry on the boundary that was recently discussed in \cite{Christensen:2013lma,Christensen:2013rfa,Moroz:2014ska,Hartong:2014oma,Hartong:2014pma,Arav:2014goa} in terms of vielbeins can also be described in the metric formalism and it would be interesting to address this in future work.

\section{The model} 

The model we consider in \cite{hvLf} is based on the generic action
\bea\label{action}
S_\x&=&\frac{1}{2\kappa^2}\int_\cm d^{d+2}x\sqrt{-g}e^{d\x \f}\times\NO\\
&&\left(R-\a_\x (\pa\f)^2-Z_\x F^2-W_\x B^{2}-V_\x\right)\NO\\
&&+\frac{1}{2\kappa^2}\int_{\pa\cm} d^{d+1}x\sqrt{-\g}2e^{d\x\f}K,
\eea
where the functions $Z_\x(\f)$, $W_\x(\f)$ and $V_\x(\f)$ are a priori arbitrary. The last term involving $K$, the trace of the extrinsic curvature, is the Gibbons-Hawking term. The parameter $\x$ defines the Weyl frame and allows us to easily translate solutions from one Weyl frame to another. This will be particularly important in order to transform the hvLf metric (\ref{thetaz}) to the dual frame.  
The $\x$ dependence of (\ref{action}) corresponds to the Weyl transformation $g\to e^{2\x\f}g$ of the $\x=0$ (Einstein frame) action. In particular, $\a_\x=\a-d(d+1)\x^2$, $Z_\x(\f)=e^{-2\x\f}Z(\f)$,
$W_\x(\f)=W(\f)$, and $V_\x(\f)=e^{2\x\f}V(\f)$,
where the quantities without the subscript $\x$ correspond to the Einstein frame. Note that the parameter $\a$ could be absorbed by a rescaling of the scalar field, but it is convenient to keep it explicitly as a parameter. Finally, 
despite the presence of a mass term for the $U(1)$ vector field, gauge invariance has been preserved by introducing the St\"uckelberg field $\om$ such that $B_\m=A_\m-\pa_\m\om$ is gauge invariant.

\section{Lifshitz \& Hyperscaling violating Lifshitz backgrounds} 

In order for the action (\ref{action}) to admit Lifshitz and hvLf solutions the scalar potential, gauge kinetic function and vector mass should respectively be of the form $ 
V_{\xi}= V_o e^{2  (\r +\xi ) \f}$,  $Z_{\xi}= Z_o e^{-2  (\xi +\n )\f}$, and $W_{\xi}=W_o e^{2 \s \f}$, for some constants $V_o$, $Z_o$, $W_o$, $\n$, $\r$, and $\s$. Lifshitz solutions then take the form 
\be\label{Lifshitz-solution} 
ds^{2}=d r^2-e^{2 z r} dt^{2}+e^{2r} d\vec{x}^2,\; B=\frac{\cq  e^{\e r}}{\e Z_o} dt,\; \f=\m r,
\ee
 with
\begin{align}
\begin{aligned} 
& \rho=-\xi,\; \n=-\x+\frac{\e-z}{\m},\;\s=\frac{z-\e}{\m},\;\cq^2=\frac12Z_o(z-1)\e,\\
&\e=\frac{(\a_\x+d^2\x^2)\m^2-d\m\x+z(z-1)}{z-1},\\
& W_o= 2 Z_o \e (d + z + d \m\x  - \e),\\
& V_o=-d (1 + \m \x) (d + z + d \m \x) - (z - 1) \e.
\end{aligned}
\end{align}
Transforming these solutions to the Einstein frame one sees that they correspond to hvLf backgrounds with $\th=-d\x\m$, which are equivalent to the solutions presented in \cite{Gouteraux:2012yr,Gath:2012pg}. Hence, (\ref{Lifshitz-solution}) describes these hvLf solutions in the dual frame. Both Lifshitz and hvLf asymptotic solutions in the Einstein frame can therefore be studied by considering only asymptotically Lifshitz solutions in a generic Weyl frame. 
Moreover, since we only demand that the action (\ref{action}) admits Lifshitz and hvLf solutions asymptotically, the above conditions on the functions $V_{\xi}$, $Z_{\xi}$ and $W_{\xi}$ are only required to hold asymptotically.

\section{Radial Hamiltonian formalism} 

The aim of the analysis carried out in \cite{hvLf} is to systematically construct the most general asymptotically locally Lifshitz solutions of the equations of motion following from the action (\ref{action}), and to use them in order to define the holographic dictionary. This analysis not only provides a concrete definition of what is meant my asymptotically locally Lifshitz solutions, but also leads to an algorithm for obtaining such solutions for the broad class of theories defined by the action (\ref{action}). Both these crucial aspects of the analysis are encoded in a `fake superpotential' which corresponds to a zero-derivative solution of the radial Hamilton-Jacobi (HJ) equation. In fact, the full holographic dictionary and the general asymptotic solutions of the equations of motion are obtained through a recursive solution of the radial Hamitlton-Jacobi equation, as we now briefly review. 

The radial Hamiltonian formalism for the action (\ref{action})
starts with a decomposition of the bulk fields as  
\bea
&&ds^2=(N^2+N_iN^i)dr^2+2N_idr dx^i+\g_{ij}dx^idx^j,\NO\\
&&A=A_r dr+A_idx^i,
\eea
where the canonical radial coordinate $r$ plays the role of Hamiltonian `time'. The Hamiltonian one obtains by inserting this decomposition in the action (\ref{action}) takes the form 
\be
H=\int d^{d+1}x\left(N\ch+N_i\ch^i+A_r\cf\right),
\ee 
where the canonical momenta of the fields $N$, $N_i$ and $A_r$ vanish identically and therefore correspond to non-dynamical Lagrange multipliers. The equations of motion for these Lagrange multipliers impose the usual Hamiltonian, momentum and $U(1)$ gauge constraints, namely
	\begin{align}
		\label{constraints}
		\begin{aligned}
			0=\ch=&-\frac{\kappa^2}{\sqrt{-\g}}e^{-d\x\f}\left\{2\left(\p^{ij}\p_{ij}-\frac1d\p^2\right)+\frac{1}{2\a}\left(\p_\f-2\x\p\right)^2\right.\\
			&\left.
			+\frac14Z^{-1}_\x(\f)\p^i\p_i+\frac12W^{-1}_\x(\f)\p_\om^2\right\}\\
	&+\frac{\sqrt{-\g}}{2\kappa^2}e^{d\x\f}\left(-R[\g]+\a_\x\pa^i\f\pa_i\f+Z_\x(\f)F^{ij}F_{ij}\right.\\
	&\left.+W_\x(\f)B^iB_i+V_\x(\f)\right),\\
    0=\ch^i=&-2D_j\p^{ji}+F^i{}_j\p^j+\p_\f\pa^i\f-B^i\p_\om,\\
	0=\cf=&-D_i\p^i+\p_\om,
		\end{aligned}
	\end{align}
where $\p^{ij}$, $\p^i$, $\p_\f$ and $\p_\om$ are the canonical momenta conjugate to the induced fields $\g_{ij}$, $A_i$, $\f$ and $\om$ respectively. These first class constraints become the radial Hamilton-Jacobi equations for the theory (\ref{action}) once the canonical momenta are expressed as gradients of 
Hamilton's principal function $\cs[\g,A,\f,\om]$ as
\be\label{HJ-momenta}
\p^{ij}=\frac{\d \cs}{\d\g_{ij}},\quad \p^i=\frac{\d\cs}{\d A_i},\quad \p_\f=\frac{\d\cs}{\d\f},\quad \p_\om=\frac{\d\cs}{\d\om},
\ee
so that (\ref{constraints}) become functional partial differential equations (PDEs) for $\cs[\g,A,\f,\om]$. The momentum constraint can be automatically solved by ensuring that $\cs$ is invariant with respect to diffeomorphisms on a slice of constant radial coordinate $r$. Moreover, the $U(1)$ gauge constraint is solved by requiring that $\cs$ is a function of the gauge-invariant field $B_i$ and not individually of $A_i$ and $\om$. The only non-trivial equation therefore is the Hamiltonian constraint $\ch=0$.

\section{Asymptotically locally Lifshitz backgrounds from a superpotential}

The UV asymptotic form of the induced fields is intimately connected with the leading asymptotic form of the solution $\cs[\g,B,\f]$ of the HJ equation, since the two asymptotic behaviors are related through the first order flow equations 
\begin{equation}
	\label{flow-eqs}
		\begin{aligned}
			&\dot\g_{ij}=\frac{4\kappa^2e^{-d\x\f}}{\sqrt{-\g}}\left(\frac{\x}{2\a}\g_{ij}\frac{\d}{\d\f}-\left(\g_{ik}\g_{jl}-\frac{\a-d\x^2}{d\a}\g_{ij}\g_{kl}\right)\frac{\d}{\d\g_{kl}}\right)\cs,\\
			&\dot A_i=-\frac{\kappa^2}{2}\frac{1}{\sqrt{-\g}}e^{-d\x\f}Z_\x^{-1}(\f)
			\g_{ij}\frac{\d}{\d A_j}\cs,\\
			&\dot\f=-\frac{\kappa^2}{\a}\frac{1}{\sqrt{-\g}}e^{-d\x\f}
			\left(\frac{\d}{\d\f}-2\x\g_{ij}\frac{\d}{\d\g_{ij}}\right)\cs,\\
			&\dot\om=-\frac{\kappa^2}{\sqrt{-\g}}e^{-d\x\f}W^{-1}_\x(\f)\frac{\d}{\d\om}\cs.
		\end{aligned}
\end{equation}  
These flow equations are obtained by identifying the canonical momenta that follow from the Lagrangian (\ref{action}) with the gradients (\ref{HJ-momenta}) of the solution $\cs[\g,B,\f]$ of the HJ equation. Given an asymptotic solution of the HJ equation, these first order flow equations can be used to derive the full Fefferman-Graham asymptotic expansions for the induced fields, without having to solve asymptotically the second order equations of motion. 

Let us first consider the leading asymptotic form of the induced fields and correspondingly the leading asymptotic solution of the HJ equation, $\cs\sub{0}$. To leading order the induced fields must contain arbitrary functions of the transverse coordinates that we will later interpret as sources of gauge invariant operators in the dual quantum field theory. This implies that $\cs\sub{0}$ must not contain transverse derivatives \cite{Papadimitriou:2010as}. Requiring additionally diffeomorphism and $U(1)$ gauge invariance determines that the most general form of $\cs\sub{0}$ is
\be\label{HJ-zero-order-solution}
\cs\sub{0}=\frac{1}{\kappa^2}\int d^{d+1}x\sqrt{-\g}U(\f,B^2),
\ee
where $U(\f,B^2)$ is a yet unspecified `superpotential'. Inserting this $\cs\sub{0}$ in the Hamiltonian constraint $\ch=0$ leads to a PDE for the superpotential $U(\f,B^2)$, which we will refer to as the `superpotential equation' \cite{hvLf}. (\ref{HJ-zero-order-solution}) with the superpotential satisfying this PDE is the most general solution of the HJ equation such that the asymptotic expansions of the induced fields contain arbitrary sources. However, a generic superpotential does not lead to Lifshitz asymptotics. Requiring that $\cs\sub{0}$ leads via (\ref{flow-eqs}) to asymptotic Lifshitz scaling for the metric imposes constraints on the asymptotic form of the superpotential $U(\f,B^2)$, which in turn determine the asymptotic form of the matter fields. In particular, the gauge-invariant vector field $B_i$ must behave asymptotically as 
\be\label{Lifshitz-constraint} 
B_i\sim B_{oi}=\sqrt{-Y_o(\f)} \;\bb n_i, 
\ee
where $\bb n_i$ is the unit normal vector to the surfaces of constant time, while $Y_o(\f):=-(z-1)/2\e Z_\x(\f)$ so that $B^2_o=Y_o(\f)$. Denoting $\f=: X$ and $B^2=:Y$, the asymptotic conditions on the superpotential and its derivatives following from the requirement of Lifshitz asymptotics are
\begin{align}
	\label{U-asymptotics}
	\begin{aligned}
		& U(X,Y_o(X))\sim e^{d\x X}\left(d(1+\m\x)+z-1\right),\\
		& U_Y(X,Y_o(X))\sim -\e e^{d\x X}Z_\x(X),\\
		& U_X(X,Y_o(X))\sim e^{d\x X}\left(-\m\a_\x+d\x(d+z)\right).
	\end{aligned} 
\end{align}   
The asymptotic conditions (\ref{Lifshitz-constraint}) and (\ref{U-asymptotics}) are the {\em definition} of what is meant by asymptotically locally Lifshitz backgrounds in \cite{hvLf}. These conditions determine through the flow equations (\ref{flow-eqs}) the leading asymptotic form of all the fields.
The subleading form of the superpotential $U(\f,B^2)$ determines the subleading terms with no transverse derivatives in the asymptotic expansions of the induced fields and is obtained by solving the superpotential equation subject to the asymptotic conditions (\ref{U-asymptotics}). The full asymptotic solution for the superpotential in general takes the form of a Taylor series in $Y-Y_o$, or equivalently $B_i-B_{oi}$. The composite scalar field $Y-Y_o$ sources a scalar operator which, depending on the various parameters defining the Lagrangian, can be relevant, marginally relevant, or irrelevant. If the dual operator is relevant then a finite number of terms in the Taylor expansion suffice to obtain an asymptotic complete integral \cite{hvLf}. When the dual operator is marginally relevant however the full Taylor expansion is required, while the source of $Y-Y_o$ must be set to zero if it sources an irrelevant operator in order to preserve the Lifshitz boundary conditions. In general, the subleading form of the superpotential, and consequently of the Fefferman-Graham expansions, depends on the subleading form of the functions $V_\x$, $Z_\x$ and $W_\x$ which specify the action. Contrary to the leading form of these functions which is determined by the requirement of Lifshitz asymptotics, their subleading form is unconstrained and we do not specify it in order to cover as a wide class of models as possible.

\section{Recursive solution of the HJ equation and Hamiltonian holographic renormalization} 

The solution $\cs\sub{0}$ in (\ref{HJ-zero-order-solution}) not only defines the leading asymptotic form of the fields, but also contains all non-derivative terms in the solution of the HJ equation. However, at subleading orders the solution of the HJ equation generically contains terms involving transverse derivatives. To determine the full asymptotic solution, including derivative terms, a systematic algorithm for solving the HJ equation recursively was developed in \cite{hvLf}. This algorithm generalizes the dilatation operator method \cite{Papadimitriou:2004ap,Papadimitriou:2011qb} to asymptotically non AdS and non scale invariant backgrounds and relies on the covariant expansion of the solution $\cs$ of the HJ equation in eigenfunctions of the two commuting operators
\bea
 &&\Hat\d:= \int d^{d+1}x\left(2\g_{ij}\frac{\d}{\d\g_{ij}}+B_i\frac{\d}{\d B_i}\right),\NO\\
 &&\d_B:=\int d^{d+1}x\left(2Y^{-1}B_iB_j\frac{\d}{\d\g_{ij}}+B_i\frac{\d}{\d B_i}\right).
\eea
It is easy to see that (\ref{HJ-zero-order-solution}) is an eigenfunction of both these operators for any superpotential $U(\f,B^2)$. In particular, $\Hat\d \cs\sub{0}= (d+1) \cs\sub{0}$,  $\d_B \cs\sub{0}= \cs\sub{0}$. The fact that these two operators commute implies that we can expand $\cs$ in simultaneous eigenfunctions of both $\Hat\d$ and $\d_B$.  
The eigenfunctions of $\Hat\d$ correspond to any local covariant quantity containing a fixed power of the induced metric $\g_{ij}$, the gauge-invariant vector $B_i$ and their covariant derivatives. However, since both $B^2$ and $\f$ are eigenfunctions of $\Hat\d$ with zero eigenvalue, generic eigenfunctions of this operator can contain arbitrary functions of $B^2$ and $\f$. Diffeomorphism covariance can then be used to show that the operator $\Hat\d$ counts transverse derivatives \cite{Chemissany:2014xpa,hvLf}. In particular, a local covariant functional $\cs\sub{2k}$ containing $2k$ derivatives is an eigenfunction of the operator $\Hat\d$ with eigenvalue $d+1-2k$, with $d+1$ being the contribution of the volume element. The eigenfunctions of the operator $\d_B$ can be constructed using the fact that $\d_B$ annihilates the projector $\s^i_j:=\d^i_j-Y^{-1}B^iB_j$, namely $\d_B\s^{ij}=0$. Eigenfunctions of $\d_B$ can therefore be obtained from an eigenfunction of $\Hat\d$, containing say $2k$ derivatives, by decomposing it into a sum of up to $k+1$ terms containing $0, 1, \ldots, k$ factors of $\s^{ij}$ \cite{hvLf}. This splits an eigenfunction $\cs\sub{2k}$ of $\Hat\d$ with $2k$ derivatives into a sum of simultaneous eigenfunctions of $\Hat\d$ and $\d_B$ with respective eigenvalues $d+1-2k$ and $1-2\ell$, where $\ell=0,1,\ldots,k$. It follows that any asymptotic solution of the HJ equation whose leading form can expressed as in (\ref{HJ-zero-order-solution}) takes the form
\be\label{covariant-expansion-s}
\cs=\sum_{k=0}^\infty\cs\sub{2k}=\sum_{k=0}^\infty\sum_{\ell=0}^k\cs\sub{2k,2\ell},
\ee
where $\Hat\d\cs\sub{2k,2\ell}=(d+1-2k)\cs\sub{2k,2\ell}$, $\d_B\cs\sub{2k,2\ell}=(1-2\ell)\cs\sub{2k,2\ell}$ and $\cs\sub{0,0}\equiv\cs\sub{0}$ is given by (\ref{HJ-zero-order-solution}). For asymptotically Lifshitz backgrounds where the asymptotic condition (\ref{Lifshitz-constraint}) holds, this graded covariant expansion acquires a clear physical significance, since the projection operator $\s_{ij}$ asymptotes to the spatial metric $\bs_{ij}:=\g_{ij}+\bb n_i\bb n_j$ and hence the operator $\d_B$ counts time derivatives. 

The covariant expansion (\ref{covariant-expansion-s}) holds for any solution of the HJ equation whose leading form can be written as in (\ref{HJ-zero-order-solution}) for some superpotential $U(\f,B^2)$. However, in order to obtain asymptotically locally Lifshitz backgrounds we need to impose the asymptotic conditions (\ref{Lifshitz-constraint}) and (\ref{U-asymptotics}). In particular, (\ref{Lifshitz-constraint}) implies that for asymptotically locally Lifshitz backgrounds every eigenfunction $\cs\sub{2k,2\ell}$ in the expansion (\ref{covariant-expansion-s}) admits a Taylor expansion in $B_i-B_{oi}$, namely   
\be\label{Taylor}
\begin{aligned}
&\cs\sub{2k,2\ell}=\int d^{d+1}x\cl\sub{2k,2\ell}\\
&=\int d^{d+1}x\cl^0_{(2k,2\ell)}[\g(x),\f(x)]\\
&+\int d^{d+1}x\int d^{d+1}x'(B_i(x')-B_{oi}(x'))\cl^{1i}_{(2k,2\ell)}[\g(x),\f(x);x']\\
&+\co\left(B-B_o\right)^2.
\end{aligned}
\ee  
The fact that the most general asymptotic solution of the HJ equation corresponding to asymptotically locally Lifshitz solutions of the action (\ref{action}) takes the form (\ref{covariant-expansion-s}) with $\cs\sub{2k,2\ell}$ of the form (\ref{Taylor}) is the main result of \cite{hvLf}.  

Inserting the expansions (\ref{covariant-expansion-s}) and (\ref{Taylor}) in the Hamiltonian constraint $\ch=0$ we obtain a set of {\em linear} recursion relations in $k,\ell$ at every order in the Taylor expansion in $B-B_o$. At zero order in the Taylor expansion these recursion relations are \cite{hvLf} 
\be\label{kth-order-0}
\ck^{-1}(\f)\left(\frac{\d}{\d\f}\int d^{d+1}{x'}
-\cc_{k,\ell}\ca'(\f)\right)\cl^0_{(2k,2\ell)}=\car^0_{(2k,2\ell)},
\ee
where $\cc_{k,\ell}=(d+1-2k)+(z-1)(1-2\ell)$, $\ca(\f)$ is given by 
\be
e^{\ca(\f)}=Z_\x^{-\frac{1}{2(\e-z)}}\sim e^{\f/\m},
\ee
and $\ck(\f)\sim 1/\m$ is a function of the scalar field determined by $V_\x(\f)$, $Z_\x(\f)$ and $W_\x(\f)$. Moreover, the inhomogeneous term $\car^0_{(2k,2\ell)}$ of this linear equation is given by the 2-derivative terms of the HJ equation for $k=1$, or by the lower order momenta for $k>1$. Similar recursion relations hold for the higher orders in the $B-B_{oi}$ expansion \cite{hvLf}. These recursion relations are linear functional differential equations in one variable -- the scalar field $\f$ -- and can be systematically solved using the technique developed in \cite{Papadimitriou:2011qb}. They constitute a general algorithm for solving the HJ equation systematically for any asymptotically locally Lifshitz solution of the action (\ref{action}). As we now briefly review, this asymptotic solution of the HJ equation can then be used to obtain the full holographic dictionary for such backgrounds.

\section{Holographic dictionary}

The form of the recursion relations (\ref{kth-order-0}) depends on the parameter $\m$ that determines the leading asymptotic behavior of the dilaton $\f$. For $\m=0$ the dilaton asymptotes to a constant and so scale invariance is restored at least in the far UV. In that case the dual theory has a Lifshitz UV fixed point, as is the case for the Einstein-Proca action \cite{Ross:2011gu,Mann:2011hg,Griffin:2011xs,Baggio:2011cp,Baggio:2011ha}, or the Einstein-Proca-Scalar model in \cite{Chemissany:2012du,Christensen:2013lma,Christensen:2013rfa}. In all cases where the dilaton is not running ($\m=0$), the recursion relations (\ref{kth-order-0}) become algebraic instead of differential equations, which renders their solution drastically simpler. The same observation holds for the recursion relations corresponding to the higher order terms in the Taylor expansion in $B-B_o$. For $\m\neq 0$ the recursion relations are linear functional differential equations in the dilaton $\f$ and can be solved using the systematic method developed in \cite{Papadimitriou:2011qb}. This case arises for hvLf backgrounds since the hyperscaling violation exponent in the Einstein frame is given by $\th=-d\m\x$. Nevertheless, in the special case when the functions $V_\x(\f)$, $Z_\x(\f)$ and $W_\x(\f)$ are exactly -- and not merely asymptotically -- exponentials the recursion relations (\ref{kth-order-0}) become algebraic even for $\m\neq 0$ \cite{hvLf}.  
	
The asymptotic solution of the HJ equation we need in order to derive the holographic dictionary is obtained by solving the recursion relations until all terms containing UV divergences have been determined. To $\co(B-B_o)^0$ in the Taylor expansion, this means solving the recursion relation (\ref{kth-order-0}) for all $k,\ell$ such that $\cc_{k,\ell}-d\m\x\leq 0$, i.e. $(d+1-2k)+(z-1)(1-2\ell)-\th\leq 0$. In general, at order $\co(B-B_o)^m$ the corresponding recursion relations must be solved as long as $(d+1-2k)+(z-1)(1-2\ell)-\th-m\D_-\leq 0$, where $\D_+:=d+z+\th-\D_-$ is the dimension of the operator dual to the composite scalar $Y-Y_o$ \cite{hvLf}. The resulting solution of the HJ equation can be written schematically as 
\be\label{HJ-solution-div}
\cs_{div}=\sum_{k,\ell,m\;|\; \cc_{k,\ell}+d\m\x-m\D_-\geq 0}\int\cdots\int(B-B_{o})^m\cs^m_{(2k,2\ell)}.
\ee
All terms in this part of the solution are local, covariant under diffeomorphisms, and UV divergent. It follows that the boundary term required in order to remove the UV divergences and render the variational problem well posed for general asymptotically locally Lifshitz solutions of the action (\ref{action}) is 
\be\label{counterterms}
\cs_{ct}:= -\cs_{div}.
\ee 
When $z$, $\th$ and $\D_-$ are such that there exist non-negative integers $k$, $0\leq \ell\leq k$ and $m$ saturating the above inequality, then the corresponding term $\cs_{(2k,2\ell)}^m$ has a pole that needs regularization. As is well known from the relativistic case, this can be done using a form of dimensional regularization at the expense of introducing explicit cut-off dependence. In the scale invariant case $\m=0$ such terms correspond to non-relativistic conformal anomalies. 
In contrast to the relativistic case, the non-relativistic conformal anomaly potentially consists of terms with different number of spatial and time derivatives, as for example in the case of the $d=z=2$ Einstein-Proca theory \cite{Baggio:2011ha, Griffin:2011xs}. In the presence of a running dilaton ($\m\neq 0$) however, such logarithmic divergences can be absorbed into the running dilaton, thus removing any explicit dependence on the cut-off \cite{hvLf}. This is in agreement with the fact that theories with a running dilaton are not scale invariant and hence the concept of a conformal anomaly is ill defined for such theories.  

The divergent part (\ref{HJ-solution-div}) of the asymptotic solution of the HJ equation is of course not the complete solution. (\ref{HJ-solution-div}) is only the local part and it is determined through the above iterative process. However, starting at the order corresponding to dilatation weight zero \cite{hvLf}, there is an independent and UV finite solution of the HJ equation, $\Hat\cs_{ren}$, which can be parameterized as     
\be\label{Sren}
\Hat\cs_{ren}=\int d^{d+1}x\left(\g_{ij}\Hat\p^{ij}+B_i\Hat\p^i+\f\Hat\p_\f\right),
\ee 
with $\Hat\p^{ij}$, $\Hat\p^i$ and $\Hat\p_\f$ being arbitrary integration functions, subject only to the momentum constraint (\ref{constraints}). The full asymptotic solution of the HJ equation, therefore takes the form
\be\label{HJ-solution-full}
\cs=\cs_{div} + \Hat\cs_{ren} +\cdots,
\ee
where the dots stand for terms that are asymptotically subleading with respect to $\Hat\cs_{ren}$ and hence do not survive when the UV cutoff is removed. Such terms depend on both the generalized coordinates $\g_{ij}$, $B_i$, $\f$ and the generalized momenta $\Hat\p^{ij}$, $\Hat\p^i$ and $\Hat\p_\f$, but crucially they do not contain any additional integration functions. Adding the boundary term (\ref{counterterms}) to the on-shell action $\cs$ one sees immediately that $\Hat\cs_{ren}$ is holographically identified with the renormalized generating function of connected correlation functions in the dual quantum field theory. Moreover, the integration functions $\Hat\p^{ij}$, $\Hat\p^i$ and $\Hat\p_\f$ are identified with the renormalized 1-point functions of the corresponding local gauge-invariant operators \cite{hvLf}.


Given the asymptotic solution (\ref{Sren}) of the HJ equation, one can easily determine the Fefferman-Graham asymptotic expansions for all fields, identify the full set of sources and 1-point functions for the dual operators, and derive the Ward identities they satisfy. Using the anisotropic decomposition 
\bea\label{decomposition}
&&\g_{ij}dx^idx^j=-(n^2-n_a n^a)dt^2+2n_adtdx^a+\s_{ab}dx^a dx^b,\NO\\
&&A_idx^i=a dt+ A_a dx^a,
\eea
where $a,b$ are spatial indices, for the generalized coordinates and inserting the asymptotic solution (\ref{Sren}) into the flow equations (\ref{flow-eqs}) the Fefferman-Graham expansions can be systematically obtained by solving order by order these first order equations.  To leading order asymptotically we obtain \cite{hvLf}
\bea\label{sources} 
&& n\sim e^{zr}n\sub{0}(x),\;
n_a\sim e^{2r}n\sub{0}_a(x),\;
\s_{ab}\sim e^{2r}\s\sub{0}_{ab}(x),\NO\\
&&\f\sim \m r+\f\sub{0}(x), \; \j\sim e^{-\D_-r}\j_-(x),\;\om\sim\om\sub{0}(x),
\eea
where $\j:=Y_o^{-1}B_o^i(B_i-B_{oi})$ and $n\sub{0}(x)$, $n\sub{0}_a(x)$, $\s\sub{0}_{ab}(x)$, $\om\sub{0}(x)$, $\f\sub{0}(x)$ and $\j_-(x)$ are arbitrary integration functions which are identified as the sources of gauge-invariant local operators. The source of the gauge field $A_i$ is not independent since it is determined in terms of all other sources according to  
\be
A_i\sim \sqrt{\frac{z-1}{2\e Z_o}}\;n\sub{0}e^{\frac{(\e-z)\f\sub{0}}{\m}}e^{\e r}\d_{it}\left(1+e^{-\D_-r}\j_-\right)+\pa_i\om\sub{0}. 
\ee
In effect the source of $A_i$ has been traded for the source $\j_-$ of the composite scalar $\j$ and the spatial vector source for $B_a$ which is set to zero by the asymptotic condition (\ref{Lifshitz-constraint}). This means that the source of $B_a$ sources an irrelevant operator relative to the Lifshitz asymptotics, and it is in agreement with the sources found in the vielbein formalism \cite{Ross:2011gu}. The source $\om\sub{0}(x)$ amounts to a pure gauge transformation and can therefore be omitted from the list of independent sources. Indeed, the $U(1)$ Ward identity implies that it does not source an independent operator. 

Using this set of independent sources we can now identify the conjugate 1-point functions using the renormalized action in the form (\ref{Sren}). Introducing the linear combinations  
\bea\label{vevs-1}
\Hat\ct^{ij}&:=&-\frac{e^{-d\x\f}}{\sqrt{-\g}}\left(2\Hat\p^{ij}+Y_o^{-1}B_o^iB_o^jB_{ok}\Hat\p^k\right),\NO\\
\Hat\co_\f &:=&\frac{e^{-d\x\f}}{\sqrt{-\g}}\left(\Hat\p_\f+(\n+\x)B_{oi}\Hat\p^i\right),\NO\\
\Hat\co_\j&:=&\frac{e^{-d\x\f}}{\sqrt{-\g}}B_{oi}\Hat\p^i, \quad \Hat\ce^i:=  \frac{e^{-d\x\f}}{\sqrt{-\g}}\sqrt{-Y_o}\bs^i_j\Hat\p^j,
\eea
of the integration functions $\Hat\p^{ij}$, $\Hat\p^i$ and $\Hat\p_\f$, we obtain the full set of source--1-point function pairs \cite{hvLf}: 
\[\begin{array}{|l|l|}
\hline \hline 
\hspace{0.7in}\mbox{1-point function}& \hspace{0.18in}\mbox{source}\quad\\
\hline
\quad\Hat\P^i_j:=\bs^i_k\bs_{jl}\Hat\ct^{kl}\sim e^{-(d+z-\th)r}\P^i_j(x)\quad & \quad \s_{(0)ab}\\
\quad\Hat\cp^i:=-\bs^i_k\bb n_l\Hat\ct^{kl}\sim e^{-(d+2-\th)r}\cp^i(x)\quad& \quad n_{(0)a}\\
\quad\Hat\ce:=-\bb n_k\bb n_l\Hat\ct^{kl}\sim e^{-(d+z-\th)r}\ce(x)\quad& \quad n_{(0)}\\
\quad\Hat\ce^i\sim e^{-(d+2z-\th)r}\ce^i(x)\quad  &\quad 0\\
\quad\Hat\co_\f\sim  e^{-(d+z-\th)r}\co_\f(x)\quad & \quad\f_{(0)}\\
\quad\Hat\co_\j\sim  e^{-\D_+r}\co_\j(x)\quad & \quad \j_-\\
\hline\hline
\end{array}\]
The modes $\P^i_j(x)$, $\cp^i(x)$, $\ce(x)$, $\ce^i(x)$ correspond respectively to the spatial stress tensor, momentum density, energy density and energy flux. These are exactly the operators comprising the energy-momentum complex of a non-relativistic field theory \cite{Ross:2011gu}. As expected from the analysis of the sources above, the energy flux is an irrelevant operator whose source is set to zero by the requirement of Lifshitz asymptotics, in agreement with \cite{Ross:2011gu}. 

We now have all the ingredients to derive the holographic Ward identities that these 1-point functions obey. Inserting the modes $\Hat\p^{ij}$, $\Hat\p^i$ and $\Hat\p_\f$ in the momentum constraint in (\ref{constraints}) and using their decomposition in terms of the 1-point functions we obtain the full set of non-relativistic diffeomorphism Ward identities
\bea
&&\bb D_j\Hat\P^i_i+\bb n^jD_j\Hat\cp_i+\Hat\co_\f\bb D_i\f+\Hat\co_\j\bb D_i\j=0,\NO\\
&&\bb n^iD_i\Hat\ce+\bb D_i\Hat\ce^i+\Hat\co_\f\bb n^iD_i\f=0,\quad \bb D_i\Hat\cp^i=0,
\eea
where $\bb D_i$ is the covariant derivative with respect to the boundary metric $\bs_{ij}$, which we have taken here to be flat. The full expressions for arbitrary sources can be found in \cite{hvLf}. Finally, the trace or conformal Ward identity follows from the transformation of the renormalized action (\ref{Sren}) under infinitesimal anisotropic Weyl transformations and takes the form 
\be
z\Hat\ce+\Hat\P^i_i+\D_-\j\Hat\co_\j=\left\{\begin{matrix}\m\Hat\co_\f,& \m\neq 0,\\
	\ca,& \m=0,\end{matrix}\right.		
\ee
where $\ca$ is the non-relativistic conformal anomaly \cite{hvLf}. Note that for $\m\neq 0$ there is no conformal anomaly, but  instead the violation of scale invariance is encoded in the 1-point function of the dilaton operator $\Hat\co_\f$.

\section{Conclusions and outlook}

In this article we have provided a brief overview of the construction of the holographic dictionary for asymptotically locally Lifshitz and hyperscaling violating Lifshitz solutions of the general Einstein-Proca-Dilaton model (\ref{action}) which was developed in \cite{Chemissany:2014xpa,hvLf}. A concrete definition of asymptotically locally Lifshitz and hyperscaling violating Lifshitz backgrounds is obtained and the analysis is carried out for arbitrary values of the dynamical exponents $z$ and $\th$, subject only to the constraints imposed by the null energy condition. The central result of \cite{Chemissany:2014xpa,hvLf} is a general algorithm for systematically solving the radial Hamilton-Jacobi equation for the model (\ref{action}), by expanding the solution covariantly in simultaneous eigenfunctions of two commuting operators. This type of expansion generalizes the covariant expansion in eigenfunctions of the dilatation operator to non-scale invariant and non-relativistic backgrounds. The full holographic dictionary, including the full set of sources and 1-point functions, the Fefferman-Graham asymptotic expansions, and the holographic Ward identities can be obtained directly from this asymptotic solution of the radial Hamilton-Jacobi equation.  

One of the core differences in our approach relative to previous studies of hyperscaling violating backgrounds in holography is the fact that we consider such backgrounds all the way to the far UV, without using some other UV completion. This fact implies that our results can be used to address a number of interesting questions. Firstly, scale invariance usually implies (relativistic or non-relativistic) conformal invariance, which in turn imposes strong constraints on the correlation functions. Scale transformations, however, are only a conformal isometry of hyperscaling violating backgrounds. It would be interesting to determine what is the signature of such a scale `covariance' in the correlation functions of the dual field theory. Moreover, our results can also be used to define the conserved charges of asymptotically Lifshitz and hvLf black holes and study their thermodynamics. Finally, it would be very interesting to compare our results with some of the recent results for Lifshitz backgrounds using the vielbein formalism, and especially the Newton-Cartan geometry on the boundary that was discussed in \cite{Christensen:2013lma,Christensen:2013rfa,Moroz:2014ska,Hartong:2014oma,Hartong:2014pma,Arav:2014goa}.

\section*{Acknowledgements} 

	This work has been funded in part by the Consejo Superior de Investigaciones Cient\'ificas and the European Social Fund under the contract JAEDOC068, the ESF Holograv Programme, the Spanish Ministry of Economy and Competitiveness under grant FPA2012-32828, Consolider-CPAN (CSD2007-00042), the Spanish MINECO's ``Centro de Excelencia Severo Ochoa'' Programme under grant SEV-2012-0249, as well as by the grant HEPHACOS-S2009/ESP1473 from the C.A. de Madrid.










\bibliography{hvLf}

\begin{thebibliography}{10}
\expandafter\ifx\csname url\endcsname\relax
  \def\url#1{\texttt{#1}}\fi
\expandafter\ifx\csname urlprefix\endcsname\relax\def\urlprefix{URL }\fi
\expandafter\ifx\csname href\endcsname\relax
  \def\href#1#2{#2} \def\path#1{#1}\fi

\bibitem{Chemissany:2014xpa}
W.~Chemissany, I.~Papadimitriou, {Generalized dilatation operator method for
  non-relativistic holography}\href {http://arxiv.org/abs/1405.3965}
  {\path{arXiv:1405.3965}}.

\bibitem{hvLf}
W.~Chemissany, I.~Papadimitriou, {Lifshitz holography: The whole shebang}\href
  {http://arxiv.org/abs/1408.0795} {\path{arXiv:1408.0795}}.

\bibitem{Kachru:2008yh}
S.~Kachru, X.~Liu, M.~Mulligan, {Gravity Duals of Lifshitz-like Fixed Points},
  Phys.Rev. D78 (2008) 106005.
\newblock \href {http://arxiv.org/abs/0808.1725} {\path{arXiv:0808.1725}},
  \href {http://dx.doi.org/10.1103/PhysRevD.78.106005}
  {\path{doi:10.1103/PhysRevD.78.106005}}.

\bibitem{Taylor:2008tg}
M.~Taylor, {Non-relativistic holography}\href {http://arxiv.org/abs/0812.0530}
  {\path{arXiv:0812.0530}}.

\bibitem{Balasubramanian:2008dm}
K.~Balasubramanian, J.~McGreevy, {Gravity duals for non-relativistic CFTs},
  Phys.Rev.Lett. 101 (2008) 061601.
\newblock \href {http://arxiv.org/abs/0804.4053} {\path{arXiv:0804.4053}},
  \href {http://dx.doi.org/10.1103/PhysRevLett.101.061601}
  {\path{doi:10.1103/PhysRevLett.101.061601}}.

\bibitem{Son:2008ye}
D.~Son, {Toward an AdS/cold atoms correspondence: A Geometric realization of
  the Schrodinger symmetry}, Phys.Rev. D78 (2008) 046003.
\newblock \href {http://arxiv.org/abs/0804.3972} {\path{arXiv:0804.3972}},
  \href {http://dx.doi.org/10.1103/PhysRevD.78.046003}
  {\path{doi:10.1103/PhysRevD.78.046003}}.

\bibitem{Charmousis:2010zz}
C.~Charmousis, B.~Gouteraux, B.~Kim, E.~Kiritsis, R.~Meyer, {Effective
  Holographic Theories for low-temperature condensed matter systems}, JHEP 1011
  (2010) 151.
\newblock \href {http://arxiv.org/abs/1005.4690} {\path{arXiv:1005.4690}},
  \href {http://dx.doi.org/10.1007/JHEP11(2010)151}
  {\path{doi:10.1007/JHEP11(2010)151}}.

\bibitem{Huijse:2011ef}
L.~Huijse, S.~Sachdev, B.~Swingle, {Hidden Fermi surfaces in compressible
  states of gauge-gravity duality}, Phys.Rev. B85 (2012) 035121.
\newblock \href {http://arxiv.org/abs/1112.0573} {\path{arXiv:1112.0573}},
  \href {http://dx.doi.org/10.1103/PhysRevB.85.035121}
  {\path{doi:10.1103/PhysRevB.85.035121}}.

\bibitem{Dong:2012se}
X.~Dong, S.~Harrison, S.~Kachru, G.~Torroba, H.~Wang, {Aspects of holography
  for theories with hyperscaling violation}, JHEP 1206 (2012) 041.
\newblock \href {http://arxiv.org/abs/1201.1905} {\path{arXiv:1201.1905}},
  \href {http://dx.doi.org/10.1007/JHEP06(2012)041}
  {\path{doi:10.1007/JHEP06(2012)041}}.

\bibitem{Shaghoulian:2011aa}
E.~Shaghoulian, {Holographic Entanglement Entropy and Fermi Surfaces}, JHEP
  1205 (2012) 065.
\newblock \href {http://arxiv.org/abs/1112.2702} {\path{arXiv:1112.2702}},
  \href {http://dx.doi.org/10.1007/JHEP05(2012)065}
  {\path{doi:10.1007/JHEP05(2012)065}}.

\bibitem{Kanitscheider:2008kd}
I.~Kanitscheider, K.~Skenderis, M.~Taylor, {Precision holography for
  non-conformal branes}, JHEP 0809 (2008) 094.
\newblock \href {http://arxiv.org/abs/0807.3324} {\path{arXiv:0807.3324}},
  \href {http://dx.doi.org/10.1088/1126-6708/2008/09/094}
  {\path{doi:10.1088/1126-6708/2008/09/094}}.

\bibitem{Wiseman:2008qa}
T.~Wiseman, B.~Withers, {Holographic renormalization for coincident Dp-branes},
  JHEP 0810 (2008) 037.
\newblock \href {http://arxiv.org/abs/0807.0755} {\path{arXiv:0807.0755}},
  \href {http://dx.doi.org/10.1088/1126-6708/2008/10/037}
  {\path{doi:10.1088/1126-6708/2008/10/037}}.

\bibitem{Gherghetta:2001iv}
T.~Gherghetta, Y.~Oz, {Supergravity, nonconformal field theories and brane
  worlds}, Phys.Rev. D65 (2002) 046001.
\newblock \href {http://arxiv.org/abs/hep-th/0106255}
  {\path{arXiv:hep-th/0106255}}, \href
  {http://dx.doi.org/10.1103/PhysRevD.65.046001}
  {\path{doi:10.1103/PhysRevD.65.046001}}.

\bibitem{Peet:1998wn}
A.~W. Peet, J.~Polchinski, {UV / IR relations in AdS dynamics}, Phys.Rev. D59
  (1999) 065011.
\newblock \href {http://arxiv.org/abs/hep-th/9809022}
  {\path{arXiv:hep-th/9809022}}, \href
  {http://dx.doi.org/10.1103/PhysRevD.59.065011}
  {\path{doi:10.1103/PhysRevD.59.065011}}.

\bibitem{Donos:2010tu}
A.~Donos, J.~P. Gauntlett, {Lifshitz Solutions of D=10 and D=11 supergravity},
  JHEP 1012 (2010) 002.
\newblock \href {http://arxiv.org/abs/1008.2062} {\path{arXiv:1008.2062}},
  \href {http://dx.doi.org/10.1007/JHEP12(2010)002}
  {\path{doi:10.1007/JHEP12(2010)002}}.

\bibitem{Gouteraux:2012yr}
B.~Gouteraux, E.~Kiritsis, {Quantum critical lines in holographic phases with
  (un)broken symmetry}\href {http://arxiv.org/abs/1212.2625}
  {\path{arXiv:1212.2625}}.

\bibitem{Gath:2012pg}
J.~Gath, J.~Hartong, R.~Monteiro, N.~A. Obers, {Holographic Models for Theories
  with Hyperscaling Violation}\href {http://arxiv.org/abs/1212.3263}
  {\path{arXiv:1212.3263}}.

\bibitem{Bueno:2012vx}
P.~Bueno, W.~Chemissany, C.~Shahbazi, {On hvLif-like solutions in gauged
  Supergravity}\href {http://arxiv.org/abs/1212.4826} {\path{arXiv:1212.4826}}.

\bibitem{Narayan:2012wn}
K.~Narayan, {AdS null deformations with inhomogeneities}\href
  {http://arxiv.org/abs/1209.4348} {\path{arXiv:1209.4348}}.

\bibitem{Moroz:2014ska}
S.~Moroz, C.~Hoyos, {Effective theory of two-dimensional chiral superfluids:
  gauge duality and Newton-Cartan formulation}\href
  {http://arxiv.org/abs/1408.5911} {\path{arXiv:1408.5911}}.

\bibitem{Hartong:2014oma}
J.~Hartong, E.~Kiritsis, N.~A. Obers, {Lifshitz Space-Times for Schroedinger
  Holography}\href {http://arxiv.org/abs/1409.1519} {\path{arXiv:1409.1519}}.

\bibitem{Hartong:2014pma}
J.~Hartong, E.~Kiritsis, N.~A. Obers, {Schroedinger Invariance from Lifshitz
  Isometries in Holography and Field Theory}\href
  {http://arxiv.org/abs/1409.1522} {\path{arXiv:1409.1522}}.

\bibitem{Arav:2014goa}
I.~Arav, S.~Chapman, Y.~Oz, {Lifshitz Scale Anomalies}\href
  {http://arxiv.org/abs/1410.5831} {\path{arXiv:1410.5831}}.

\bibitem{Ross:2009ar}
S.~F. Ross, O.~Saremi, {Holographic stress tensor for non-relativistic
  theories}, JHEP 0909 (2009) 009.
\newblock \href {http://arxiv.org/abs/0907.1846} {\path{arXiv:0907.1846}},
  \href {http://dx.doi.org/10.1088/1126-6708/2009/09/009}
  {\path{doi:10.1088/1126-6708/2009/09/009}}.

\bibitem{Horava:2009vy}
P.~Horava, C.~M. Melby-Thompson, {Anisotropic Conformal Infinity},
  Gen.Rel.Grav. 43 (2011) 1391--1400.
\newblock \href {http://arxiv.org/abs/0909.3841} {\path{arXiv:0909.3841}},
  \href {http://dx.doi.org/10.1007/s10714-010-1117-y}
  {\path{doi:10.1007/s10714-010-1117-y}}.

\bibitem{Ross:2011gu}
S.~F. Ross, {Holography for asymptotically locally Lifshitz spacetimes},
  Class.Quant.Grav. 28 (2011) 215019.
\newblock \href {http://arxiv.org/abs/1107.4451} {\path{arXiv:1107.4451}},
  \href {http://dx.doi.org/10.1088/0264-9381/28/21/215019}
  {\path{doi:10.1088/0264-9381/28/21/215019}}.

\bibitem{Mann:2011hg}
R.~B. Mann, R.~McNees, {Holographic Renormalization for Asymptotically Lifshitz
  Spacetimes}, JHEP 1110 (2011) 129.
\newblock \href {http://arxiv.org/abs/1107.5792} {\path{arXiv:1107.5792}},
  \href {http://dx.doi.org/10.1007/JHEP10(2011)129}
  {\path{doi:10.1007/JHEP10(2011)129}}.

\bibitem{Baggio:2011cp}
M.~Baggio, J.~de~Boer, K.~Holsheimer, {Hamilton-Jacobi Renormalization for
  Lifshitz Spacetime}, JHEP 1201 (2012) 058.
\newblock \href {http://arxiv.org/abs/1107.5562} {\path{arXiv:1107.5562}},
  \href {http://dx.doi.org/10.1007/JHEP01(2012)058}
  {\path{doi:10.1007/JHEP01(2012)058}}.

\bibitem{Baggio:2011ha}
M.~Baggio, J.~de~Boer, K.~Holsheimer, {Anomalous Breaking of Anisotropic
  Scaling Symmetry in the Quantum Lifshitz Model}, JHEP 1207 (2012) 099.
\newblock \href {http://arxiv.org/abs/1112.6416} {\path{arXiv:1112.6416}},
  \href {http://dx.doi.org/10.1007/JHEP07(2012)099}
  {\path{doi:10.1007/JHEP07(2012)099}}.

\bibitem{Griffin:2011xs}
T.~Griffin, P.~Horava, C.~M. Melby-Thompson, {Conformal Lifshitz Gravity from
  Holography}, JHEP 1205 (2012) 010.
\newblock \href {http://arxiv.org/abs/1112.5660} {\path{arXiv:1112.5660}},
  \href {http://dx.doi.org/10.1007/JHEP05(2012)010}
  {\path{doi:10.1007/JHEP05(2012)010}}.

\bibitem{Griffin:2012qx}
T.~Griffin, P.~Horava, C.~M. Melby-Thompson, {Lifshitz Gravity for Lifshitz
  Holography}\href {http://arxiv.org/abs/1211.4872} {\path{arXiv:1211.4872}}.

\bibitem{Cassani:2011sv}
D.~Cassani, A.~F. Faedo, {Constructing Lifshitz solutions from AdS}, JHEP 1105
  (2011) 013.
\newblock \href {http://arxiv.org/abs/1102.5344} {\path{arXiv:1102.5344}},
  \href {http://dx.doi.org/10.1007/JHEP05(2011)013}
  {\path{doi:10.1007/JHEP05(2011)013}}.

\bibitem{Chemissany:2011mb}
W.~Chemissany, J.~Hartong, {From D3-Branes to Lifshitz Space-Times},
  Class.Quant.Grav. 28 (2011) 195011.
\newblock \href {http://arxiv.org/abs/1105.0612} {\path{arXiv:1105.0612}},
  \href {http://dx.doi.org/10.1088/0264-9381/28/19/195011}
  {\path{doi:10.1088/0264-9381/28/19/195011}}.

\bibitem{Chemissany:2012du}
W.~Chemissany, D.~Geissbuhler, J.~Hartong, B.~Rollier, {Holographic
  Renormalization for z=2 Lifshitz Space-Times from AdS}\href
  {http://arxiv.org/abs/1205.5777} {\path{arXiv:1205.5777}}.

\bibitem{Christensen:2013lma}
M.~H. Christensen, J.~Hartong, N.~A. Obers, B.~Rollier, {Torsional
  Newton-Cartan Geometry and Lifshitz Holography}, Phys.Rev. D89 (2014) 061901.
\newblock \href {http://arxiv.org/abs/1311.4794} {\path{arXiv:1311.4794}},
  \href {http://dx.doi.org/10.1103/PhysRevD.89.061901}
  {\path{doi:10.1103/PhysRevD.89.061901}}.

\bibitem{Christensen:2013rfa}
M.~H. Christensen, J.~Hartong, N.~A. Obers, B.~Rollier, {Boundary Stress-Energy
  Tensor and Newton-Cartan Geometry in Lifshitz Holography}, JHEP 1401 (2014)
  057.
\newblock \href {http://arxiv.org/abs/1311.6471} {\path{arXiv:1311.6471}},
  \href {http://dx.doi.org/10.1007/JHEP01(2014)057}
  {\path{doi:10.1007/JHEP01(2014)057}}.

\bibitem{Korovin:2013bua}
Y.~Korovin, K.~Skenderis, M.~Taylor, {Lifshitz as a deformation of Anti-de
  Sitter}\href {http://arxiv.org/abs/1304.7776} {\path{arXiv:1304.7776}}.

\bibitem{Korovin:2013nha}
Y.~Korovin, K.~Skenderis, M.~Taylor, {Lifshitz from AdS at finite temperature
  and top down models}\href {http://arxiv.org/abs/1306.3344}
  {\path{arXiv:1306.3344}}.

\bibitem{Papadimitriou:2010as}
I.~Papadimitriou, {Holographic renormalization as a canonical transformation},
  JHEP 1011 (2010) 014.
\newblock \href {http://arxiv.org/abs/1007.4592} {\path{arXiv:1007.4592}},
  \href {http://dx.doi.org/10.1007/JHEP11(2010)014}
  {\path{doi:10.1007/JHEP11(2010)014}}.

\bibitem{Papadimitriou:2004ap}
I.~Papadimitriou, K.~Skenderis, {AdS / CFT correspondence and geometry} (2004)
  73--101\href {http://arxiv.org/abs/hep-th/0404176}
  {\path{arXiv:hep-th/0404176}}.

\bibitem{Papadimitriou:2011qb}
I.~Papadimitriou, {Holographic Renormalization of general dilaton-axion
  gravity}, JHEP 1108 (2011) 119.
\newblock \href {http://arxiv.org/abs/1106.4826} {\path{arXiv:1106.4826}},
  \href {http://dx.doi.org/10.1007/JHEP08(2011)119}
  {\path{doi:10.1007/JHEP08(2011)119}}.

\end{thebibliography}
\bibliographystyle{elsarticle-num}
\end{document}